\documentclass[10pt,twoside]{article}
\usepackage[applemac]{inputenc}
\usepackage{amssymb,amsmath}
\def\volnum{Manuscrit}

\usepackage{aflbcours}
\pdebut{1}
\def\pacs#1{\LP P.A.C.S.: #1}
\title{ Gauge Invariance in Classical 
Electrodynamics\footnotemark[1]} 
\footnotetext[1]{This paper 
is written by W. Engelhardt in his private 
capacity. 
No official 
sup-\\port 
by the 
Max-Planck-Institut f\"ur Plasmaphysik 
is intended or should be inferred. }
\titleshort{ Gauge Invariance }

\author{ Wolfgang Engelhardt\footnotemark[2]} 
\footnotetext[2]{Home address: Fasaneriestrasse 8, D-80636 M\"unchen, Germany 
\\ 
\indent\hspace{1.4mm} 
E-mail address: wolfgangw.engelhardt@t-online.de }

\authorshort{ W. Engelhardt }
\address{ Max-Planck-Institut f\"ur Plasmaphysik, D-85741 
Garching, Germany }
\addtocounter{footnote}{2} 
\begin{document}
\maketitle

\vskip 1cm
\begin{abstract}
R\'ESUM\'E. Le concept de l'invariance de jauge dans l'\'electrodynamique 
classique suppose tacitement que les \'equations de Maxwell poss\`edent des 
solutions uniques. Mais en calculant le champ \'electromagn\'etique d'une 
particule en mouvement faisant usage de la jauge de Lorenz ainsi que de 
la jauge de Coulomb, et r\'esolvant directement les \'equations des champs 
nous obtenons des solutions contradictoires. Nous concluons donc que 
l'hypoth\`ese tacite de l'unicit\'e de la solution n'est pas justifi\'ee. 
La raison pour cette difficult\'e peut \^etre attribu\'ee aux \'equations 
d'onde inhomog\`enes qui connectent simultan\'ement les champs propageants 
et leurs sources. 

{\it 
ABSTRACT. The concept of gauge invariance in classical electrodynamics 
assumes tacitly that Maxwell's equations have unique solutions. 
By calculating the electromagnetic field of a moving particle both 
in Lorenz and in Coulomb gauge and directly from the field equations 
we obtain, however, contradicting solutions. We conclude that the 
tacit assumption of uniqueness is not justified. The reason for this 
failure is traced back to the inhomogeneous wave equations which connect 
the propagating fields and their sources at the same time.}
\end{abstract}

\newpage 

\section{ Introduction }

``The principle of gauge invariance plays a key role 
in the standard model, which describes electroweak and strong interactions 
of elementary particles.'' This statement is quoted from an 
article by J. D. Jackson and L. B. Okun [1] which 
emphasizes the importance of the principle and delineates its historical 
evolution. The concept emerged from classical electrodynamics where 
the electromagnetic field expressed in terms of potentials: 
$\vec {E} 
\, = 
\, - 
\nabla 
\phi - 
\partial 
\vec {A}/c 
\, 
\partial t, 
\; 
\vec {B}=rot 
\, 
\vec {A}$ 
does not change under the transformation: $\vec {A} 
\rightarrow 
\vec {A}+ 
\nabla 
\chi , 
\; 
\phi 
\rightarrow 
\phi - 
\partial 
\chi /c 
\, 
\partial t$. Since $div 
\, 
\vec {A} 
\rightarrow div 
\, 
\vec {A}+ 
\Delta 
\chi$ 
and $\chi$ is an arbitrary function, 
the divergence of the vector potential can seemingly be chosen arbitrarily 
without influencing the fields. This feature was exploited to de-couple 
Maxwell's inhomogeneous equations by either choosing $div 
\, 
\vec {A}+ 
\partial 
\phi /c 
\, 
\partial t=0$ 
(Lorenz\footnote{In [1] it is pointed out that Ludwig Valentin Lorenz 
published more than 
25 years before Hendrik Antoon Lorentz what is commonly known as the 
``Lorentz condition''. To give proper credit to Lorenz we use in 
this paper the term ``Lorenz gauge''.} gauge) or $div 
\, 
\vec {A}=0$ 
(Coulomb gauge). The solution for the fields should be entirely 
independent of this choice. 

There is, however, a tacit assumption behind the formalism 
of electrodynamic gauge invariance: Maxwell's equations must 
have unique solutions, otherwise it is meaningless to talk about potentials 
from which fields may be derived. In reference [1] 
it is said: ``It took almost a century to formulate this nonuniqueness 
of potentials that exists despite the uniqueness of the electromagnetic 
fields.'' To our knowledge it was never attempted to prove that 
the electromagnetic field resulting from a solution of Maxwell's 
equations is actually unique, it was just taken for granted. If there 
were no unique solution of Maxwell's linear system of first 
order field equations, gauge transformations on (nonexisting) potentials 
would be irrelevant. 

In this paper we start with the usual assumption that unique 
solutions of Maxwell's equations do exist and try to calculate 
them both with the help of potentials (Sections 2 - 4) and 
directly from the field equations (Section 5). For the electromagnetic 
field of a moving particle we find, however, contradicting solutions 
depending on the method used. In Section 6 we show that the standard 
Li\'enard-Wiechert fields, which satisfy Maxwell's source-free 
equations, cannot be considered as a unique solution of Maxwell's 
inhomogeneous equations. Thus, we infer that the tacit assumption 
concerning the existence of unique solutions is not justified in general. 

The reason for this failure is discussed in Section 8 where 
we come to the conclusion that the mixture of elliptic and hyperbolic 
equations, as formulated by Maxwell, does not permit a physical solution 
for moving point sources. 

\vskip 30pt 
\section{The electromagnetic field of a moving particle calculated 
in Lorenz gauge} 

The electromagnetic field produced by a moving particle is 
calculated [2] from Maxwell's equations: 
\begin{equation} 
div\, 
\vec {E} 
\, = 
\, 4 
\, 
\pi 
\, 
\rho 
\end{equation} 
\begin{equation} 
rot 
\, 
\vec {E} 
\, = 
\, -{{1} \over {c}} 
\, {{ 
\partial 
\vec {B}} \over { 
\partial t}} 
\end{equation} 
\begin{equation} 
div\, 
\vec {B} 
\, = 
\, 0 
\end{equation} 
\begin{equation} 
rot 
\, 
\vec {B}={{4 
\, 
\pi } \over {c}} 
\, 
\rho 
\, 
\vec {v} 
\, + 
\, {{1} \over {c}} 
\, {{ 
\partial 
\vec {E}} \over { 
\partial t}} 
\end{equation} 
where $\rho$ is the charge density located in a narrow region 
round a center moving with velocity $\vec {v}$. The total 
charge of the particle is the integral over all space: 
\begin{equation} 
\, 
\int \limits 
\!
\! 
\int \limits 
\! 
\! 
\int \limits 
\rho 
\, 
\left( { 
\vec {x} 
\, ', 
\, t} 
\right) 
\, 
\, d^{3}x'=e 
\end{equation} 
The usual method to solve equations (1-4) is to adopt the 
potential ansatz: 
\begin{equation} 
\vec {E} 
\, = 
\, 
\, -{{1} \over {c}} 
\, {{ 
\partial 
\vec {A}} \over { 
\partial t}} 
\, - 
\, 
\nabla 
\phi 
\; 
\; , 
\; 
\; 
\vec {B} 
\, = 
\, rot 
\, 
\vec {A} 
\end{equation} 
which satisfies automatically equations (2) and (3) and leads to two 
second order differential equations: 
\begin{equation} 
\Delta 
\phi +{{1} \over {c}} 
\, div 
\, {{ 
\partial 
\vec {A}} \over { 
\partial t}} 
\, = 
\, 
\, - 
\, 
\! 4 
\, 
\pi 
\, 
\rho 
\end{equation} 
\begin{equation} 
\Delta 
\vec {A}-{{1} \over {c^{2}}} 
\, {{ 
\partial ^{2} 
\vec {A}} \over { 
\partial t^{2}}}= 
\, -{{4 
\, 
\pi } \over {c}} 
\, 
\rho 
\, 
\vec {v}+ 
\nabla 
\left( {div 
\, 
\vec {A}+{{1} \over {c}} 
\, {{ 
\partial 
\phi } \over { 
\partial t}}} 
\right) 
\end{equation} 

The freedom in the choice of the divergence of the vector potential 
is exploited to de-couple equations (7) and (8). Adopting the Lorenz 
condition: 
\begin{equation}div 
\, 
\vec {A}+{{1} \over {c}} 
\, {{ 
\partial 
\phi } \over { 
\partial t}}=0 
\end{equation} 
one obtains two wave equations of the same structure: 
\begin{equation} 
\Delta 
\phi -{{1} \over {c^{2}}} 
\, {{ 
\partial ^{2} 
\phi } \over { 
\partial t^{2}}}= 
\, 
\, - 
\, 4 
\, 
\pi 
\, 
\rho 
\end{equation} 
\begin{equation} 
\Delta 
\vec {A}-{{1} \over {c^{2}}} 
\, {{ 
\partial ^{2} 
\vec {A}} \over { 
\partial t^{2}}}= 
\, - 
\! 
\, {{4 
\, 
\pi } \over {c}} 
\rho 
\, 
\vec {v} 
\end{equation} 
The solution of (10) subject to the boundary condition that the scalar 
potential vanishes at infinity is: 
\begin{equation} 
\phi 
\, 
\left( { 
\vec {x} 
\, ,t} 
\right) 
\, = 
\, 
\, 
\int \limits 
\! 
\! 
\int \limits 
\! 
\! 
\int \limits {{ 
\rho 
\, 
\left( { 
\vec {x} 
\, ', 
\, t-{{R} \over {c}}} 
\right) } \over {R}} 
\, d^{3}x' 
\; 
\; , 
\; 
\; 
\; R= 
\left\vert { 
\vec {x}- 
\vec {x} 
\, '} 
\right\vert 
\end{equation} 
where the position of the charge density is to be taken at the retarded 
time: 
\begin{equation} 
t'=t-{{R} \over {c}} 
\end{equation} 
The advanced solution with $t'=t+R/c$ is excluded on physical 
grounds. 

The integral (12) may be carried out by employing the $\delta$ 
- function formalism: 
\begin{equation} 
\rho 
\, 
\left( { 
\vec {x} 
\, ', 
\, t-{{R} \over {c}}} 
\right) 
\, = 
\, 
\, 
\int \limits _{- 
\infty }^{+ 
\infty } 
\! 
\rho 
\, 
\left( { 
\vec {x} 
\, ', 
\, t'} 
\right) 
\; 
\, 
\delta 
\, 
\left( {t'-t+{{R} \over {c}}} 
\right) 
\, dt' 
\end{equation} 
Substituting this into (12) yields: 
\begin{equation} 
\phi 
\, 
\left( { 
\vec {x} 
\, ,t} 
\right) 
\, = 
\, 
\, 
\int \limits _{- 
\infty }^{+ 
\infty } 
\int \limits 
\! 
\! 
\int \limits 
\! 
\! 
\int \limits {{ 
\rho 
\, 
\left( { 
\vec {x} 
\, ', 
\, t'} 
\right) } \over {R}} 
\; 
\delta 
\, 
\left( {t'-t+{{R} \over {c}}} 
\right) 
\, d^{3}x' 
\, dt' 
\end{equation} 
The integration over all space for a point-like particle results with 
(5) in: 
\begin{equation} 
\phi 
\, 
\left( { 
\vec {x} 
\, ,t} 
\right) 
\, = 
\, 
\, 
\int \limits _{- 
\infty }^{+ 
\infty }{{e} \over {R}} 
\; 
\delta 
\, 
\left( {t'-t+{{R} \over {c}}} 
\right) 
\, 
\, dt' 
\end{equation} 
where $R$ expresses now the distance between the 
field point and the position of the charge $\vec {x} 
\, ' 
\left( {t'} 
\right)$ at the retarded time. Changing to the variable 
\begin{equation} 
u=t'-t+{{R 
\, 
\left( { 
\vec {x}, 
\, 
\vec {x} 
\, ' 
\left( {t'} 
\right) } 
\right) } \over {c}} 
\; 
\; , 
\; 
\; 
\; 
\; 
\> {{du} \over {dt'}}=1-{{ 
\vec {R}} \over { 
\, c 
\, R}} 
\cdot {{d 
\vec {x} 
\, ' 
\left( {t'} 
\right) } \over {dt'}} 
\; \; , \; \; \; {{d \vec {x} \, ' \left( {t'} 
\right) } \over {dt'}}= 
\vec {v}
\, 
\left( {t'} 
\right) 
\end{equation} 
we may integrate (16) and obtain the result: 
\begin{equation} 
\phi 
\, 
\left( { 
\vec {x} 
\, ,t} 
\right) 
\, = 
\, 
\, 
\left[ {{{e} \over {R 
\, 
\left( {1-{{ 
\vec {R} 
\cdot 
\vec {v}} \over {R 
\, c}}} 
\right) }}} 
\right] _{t'=t-{{R} \over {c}}} 
\end{equation} 
Similarly, we find from (11): 
\begin{equation} 
\vec {A} 
\, 
\left( { 
\vec {x} 
\, ,t} 
\right) 
\, = 
\, 
\, 
\left[ {{{e 
\, 
\vec {v}} \over {c 
\, R 
\, 
\left( {1-{{ 
\vec {R} 
\cdot 
\vec {v}} \over {R 
\, c}}} 
\right) }}} 
\right] _{t'=t-{{R} \over {c}}} 
\end{equation} 

Solutions (18) and (19) are the well-known retarded Li\'enard-Wiechert 
potentials. With the differentiation rules resulting from (13): 
\begin{equation} 
{{ 
\partial } \over { 
\partial t}}= {{1} \over { 
\, 1-{{ 
\vec {R} 
\cdot 
\vec {v}} \over {R 
\, c}}}} 
\, {{ 
\partial } \over { 
\partial t'}}={{1} \over { 
\lambda }} 
\, {{ 
\partial } \over { 
\partial t'}} 
\; 
\; , 
\; 
\; 
\; {{ 
\partial } \over { 
\partial 
\vec {x}}} 
\, 
\left( {f 
\, 
\left( {t'} 
\right) } 
\right) = 
-{{df} \over {dt'}} 
\, {{1} \over {c 
\, }} 
\, {{ 
\partial R} \over { 
\partial 
\vec {x}}} 
\; 
\; , 
\; 
\; 
\; {{ 
\partial R} \over { 
\partial 
\vec {x}}}={{ 
\vec {R}} \over { 
\lambda 
\, R}} 
\end{equation} 
one obtains with (6) for the fields: 
\begin{eqnarray} 
\vec {E} 
\left( { 
\vec {x} 
,t} 
\right) 
&\, = \,& e 
\left[ {{{1} \over { 
\lambda ^{3}}} 
\left( {{{ 
\vec {R}} \over {R^{3} 
\, }}-{{ 
\vec {v}} \over { 
\, c 
\, R^{2}}}} 
\right) 
\! 
\! 
\, 
\left( { 
\! 1-{{v^{2}} \over {c^{2}}} 
\, +{{1} \over {c^{2}}} 
\, 
\vec {R} 
\cdot {{d 
\vec {v}} \over {dt'}}} 
\right) - 
\, 
\! {{ 
\, 1} \over {c^{2} 
\, R 
\, 
\lambda ^{2}}} 
\, {{d 
\vec {v}} \over {dt'}}} 
\right] \nonumber \\ 
\vec {B} 
\left( { 
\vec {x} 
\, ,t} 
\right) & \, = \,& -e 
\left[ {{{ 
\, 
\vec {R} 
\times 
\vec {v}} \over { 
\, 
\! c 
\, R^{3} 
\, 
\lambda ^{3}}} 
\, 
\left( { 
\! 1-{{v^{2}} \over {c^{2}}} 
\, +{{1} \over {c^{2}}} 
\, 
\vec {R} 
\cdot {{d 
\vec {v}} \over {dt'}}} 
\right) +{{ 
\, 
\vec {R}} \over { 
\, 
\! c^{2} 
\, R^{2} 
\, 
\lambda ^{2}}} 
\times {{d 
\vec {v}} \over {dt'}}} 
\right] 
\end{eqnarray} 
where $ t'=t-{{R} \over {c} } $.

\vskip 30pt 
\section{Solution in Coulomb gauge} 

The fields as given by (21) must be the same when they are 
calculated in Coulomb gauge by adopting the condition: 
\begin{equation} 
div 
\, 
\vec {A}_{}=0 
\end{equation} 
Equations (7) and (8) become: 
\begin{equation} 
\Delta 
\phi = 
\, 
\, - 
\, 4 
\, 
\pi 
\, 
\rho 
\end{equation} 
\begin{equation} 
\Delta 
\vec {A}-{{1} \over {c^{2}}} 
\, {{ 
\partial ^{2} 
\vec {A}} \over { 
\partial t^{2}}}= 
\, - 
\! 
\, {{4 
\, 
\pi } \over {c}} 
\rho 
\, 
\vec {v}+{{1} \over {c}} 
\, {{ 
\partial } \over { 
\partial t}} 
\nabla 
\phi 
\end{equation} 
Solution of (23) yields the instantaneous Coulomb potential: 
\begin{equation} 
\phi _{C} 
\, 
\left( { 
\vec {x} 
\, ,t} 
\right) 
\, = 
\, 
\, {{e 
\, } \over { 
\, r}} 
\; 
\; , 
\; 
\; 
\; r= 
\left\vert { 
\vec {x}- 
\vec {x} 
\, ' 
\left( {t} 
\right) 
\, } 
\right\vert 
\end{equation} 
which substituted into (24) results in a wave equation for the `Coulomb' 
vector potential: 
\begin{equation} 
\Delta 
\vec {A}_{C}-{{1} \over {c^{2}}} 
\, {{ 
\partial ^{2} 
\vec {A}_{C}} \over { 
\partial t^{2}}}= 
\, - 
\! 
\, {{4 
\, 
\pi } \over {c}} 
\rho 
\, 
\vec {v}-{{e} \over {c}} 
\, {{ 
\partial } \over { 
\partial t}} 
\, 
\left( {{{ 
\vec {r}} \over {r^{3}}}} 
\right) 
\end{equation} 
The vector $ 
\vec {r}= 
\vec {x}- 
\vec {x}\,' 
\left( {t} 
\right)$ 
denotes now the simultaneous distance 
between charge and field point. The first term on the right-hand-side 
of (26) yields a contribution to the vector potential which is identical 
with (19): 
\begin{equation} 
\vec {A}_{C1} 
\, 
\left( { 
\vec {x} 
\, ,t} 
\right) 
\, = 
\, 
\, 
\left[ {{{e 
\, 
\vec {v}} \over {c 
\, R 
\, 
\left( {1-{{ 
\vec {R} 
\cdot 
\vec {v}} \over {R 
\, c}}} 
\right) }}} 
\right] _{t'=t-{{R} \over {c}}} 
\end{equation} 
The solution of the wave equation for the second part of the vector 
potential may be written in the form: 
\begin{eqnarray} 
\vec {A}_{C2} 
\, 
\left( { 
\vec {x} 
\, ,t} 
\right) &\, =\, &{{e} \over {4 
\, 
\pi 
\, c}} 
\int \limits _{- 
\infty }^{+ 
\infty } 
\int \limits 
\! 
\! 
\int \limits 
\! 
\! 
\int \limits {{ 
\partial } \over { 
\partial t'}} 
\left( {{{ 
\vec {r}} \over {r^{3}}}} 
\right) 
\, 
\delta 
\left( {t'-t+{{R} \over {c}}} 
\right) 
\, {{d^{3}s} \over {R}} 
\, dt' \nonumber \\ 
\vec {r}&\, =\, & 
\vec {s}- 
\vec {x} 
\, ' 
\left( {t'} 
\right) 
\; 
\; , 
\; 
\; 
\; R= 
\left\vert { 
\vec {s}- 
\vec {x} 
\, } 
\right\vert 
\end{eqnarray} 
when we employ a $\delta$ - function as in the previous 
Section. Here we have used the integration variable $ 
\vec {s}$ 
in distinction of the position $ 
\vec {x}\,'\left({t}\right)$ 
of the charge. The contribution of $ 
\vec {A}_{C2}$ to the electric field is with (6): 
\begin{equation} 
\vec {E}_{2} 
\, 
\left( { 
\vec {x} 
\, ,t} 
\right) = 
\, -{{e} \over {4 
\, 
\pi 
\, c^{2}}} 
\int \limits 
\! 
\! 
\int \limits 
\! 
\! 
\int \limits 
\left[ {{{ 
\partial ^{2}} \over { 
\partial t'^{2} 
\, }} 
\, 
\left( {{{ 
\vec {r}} \over {r^{3}}}} 
\right) } 
\right] _{t'=t-{{R} \over {c}}} 
\, {{d^{3}s} \over {R}} 
\end{equation} 
and the contribution to the magnetic field may be written as: 
\begin{eqnarray} 
\vec {B}_{2} 
\, 
\left( { 
\vec {x} 
\, ,t} 
\right) &\,=\,&{{e} \over {4 
\, 
\pi 
\, c}} 
\int \limits _{- 
\infty }^{+ 
\infty } 
\int \limits 
\! 
\! 
\int \limits 
\! 
\! 
\int \limits 
\nabla 
\, 
\left( {{{ 
\delta } \over {R}}} 
\right) 
\times {{ 
\partial } \over { 
\partial t'}} 
\, 
\left( {{{ 
\vec {r}} \over {r^{3}}}} 
\right) 
\, d^{3}s 
\, dt' \nonumber \\ 
&=\,&{{e} \over {4 
\, 
\pi 
\, c}} 
\int \limits _{- 
\infty }^{+ 
\infty } 
\int \limits 
\! 
\! 
\int \limits 
\! 
\! 
\int \limits 
\, 
\left( {{{ 
\delta '} \over {c 
\, R^{2}}}-{{ 
\delta } \over {R^{3}}}} 
\right) 
\, 
\vec {R} 
\times {{ 
\partial } \over { 
\partial t'}} 
\, 
\left( {{{ 
\vec {r}} \over {r^{3}}}} 
\right) 
\, d^{3}s 
\, dt' 
\end{eqnarray} 
Performing a partial integration over $t'$ we obtain 
the solution in the form: 
\begin{equation} 
\vec {B}_{2} 
\, 
\left( { 
\vec {x} 
,t} 
\right)\, = 
-{{e} \over {4 
\, 
\pi 
\, c}} 
\int \limits 
\! 
\! 
\int \limits 
\! 
\! 
\int \limits 
\, 
\left[ {{{ 
\vec {R}} \over {R^{3}}} 
\times 
\left( {{{ 
\partial } \over { 
\partial t'}} 
\, 
\left( {{{ 
\vec {r}} \over {r^{3}}}} 
\right) +{{R} \over {c}} 
\, {{ 
\partial ^{2}} \over { 
\partial t'^{ 
\, 2}}} 
\left( {{{ 
\vec {r}} \over {r^{3}}}} 
\right) } 
\right) } 
\right] _{t'=t-{{R} \over {c}}}\!\!\!\!\!\!\!\!\!\!d^{3}s 
\end{equation} 

Since (27) yields already the magnetic field as given by (21), the 
contribution (31) must vanish which is not likely to occur: The finite 
cross-product is to be integrated with different weight, so that both 
terms in (31) cannot vanish simultaneously. Furthermore, the condition 
for identical electric fields in Coulomb and in Lorenz gauge: 
\begin{equation} 
\nabla 
\phi _{LW}= 
\nabla 
\phi _{C}+{{1} \over {c}} 
\, {{ 
\partial 
\vec {A}_{C2}} \over { 
\partial t}} 
\end{equation} 
where $\phi _{LW}$ denotes the potential (18), cannot be 
satisfied, if (28) is not an irrotational field. Even if the second 
term in (32) could be written as a gradient: $\partial 
\vec {A}_{C2}/c 
\, 
\partial t= 
\nabla 
\phi _{2}$, the condition: $ 
\phi _{LW}= 
\phi _{C}+ 
\phi _{2}+const$ would still be 
violated, since $\partial 
\vec {A}_{C2}/ 
\partial t $, and thereby $\phi _{2}$, would depend on the acceleration 
according to (28), which 
is not the case for $ 
\phi _{LW}$ 
nor $ 
\phi _{C}$. 

In order to quantify these qualitative considerations the integral 
(28): 
\begin{eqnarray} 
\vec {A}_{C2} 
\, 
\left( { 
\vec {x} 
\, ,t} 
\right)&\, =& 
\, -{{e} \over {4 
\, 
\pi 
\, c}} 
\int \limits 
\! 
\! 
\int \limits 
\! 
\! 
\int \limits 
\left[ {{{ 
\vec {v}} \over {r^{3}}}-{{3 
\, 
\vec {r} 
\, 
\, 
\left( { 
\vec {r} 
\cdot 
\vec {v}} 
\right) } \over {r^{5}}}} 
\right] _{t'=t-{{R} \over {c}}} 
\, 
\, {{d^{3}s} \over {R}} \nonumber \\ 
\vec {r}&\,=\,& 
\vec {s}- 
\vec {x} 
\, ' 
\left( {t'} 
\right) 
\; 
\; , 
\; 
\; 
\; 
\vec {R}\,=\, 
\vec {s}- 
\vec {x} 
\end{eqnarray} 
may be evaluated analytically for the case of a constant velocity 
of the charge: 
\begin{equation} 
\vec {x} 
\, ' 
\left( {t'} 
\right) = 
\vec {x}_{0}+ 
\vec {v}_{0} 
\, t'= 
\vec {x}_{0}+ 
\vec {v}_{0} 
\, 
\left( {t-{{R} \over {c}}} 
\right) 
\end{equation} 
The integration variable $ 
\vec {s}$ may be replaced by $\vec {R}$ so that the vector 
$\vec {r}$ may be written as: 
\begin{equation} 
\vec {r}= 
\vec {R}+ 
\vec {x}- 
\left( { 
\vec {x}_{0}+ 
\vec {v}_{0} 
\, t} 
\right) +{{R} \over {c}} 
\, 
\vec {v}_{0} 
\end{equation} 
We assume that the charge moves along the z-axis of a coordinate system 
having its origin at $ 
\vec {R}=0$. The z-component of the 
vector potential evaluated on the z-axis becomes then: 
\begin{eqnarray} 
A_{C2z}&=& 
\, -{{e 
\, v_{z}} \over {4 
\, 
\pi 
\, c}} 
\int \limits 
\! 
\! 
\int \limits 
\! 
\! 
\int \limits 
\left[ {{{1} \over {r^{3}}}-{{3 
\, 
\, 
\left( {R_{z}+ 
\beta 
\, R+z} 
\right) ^{2}} \over {r^{5}}}} 
\right] 
\, 
\, {{d^{3}R} \over {R}} 
\; 
\; , 
\; 
\; 
\; 
\beta ={{v_{z}} \over {c}} \nonumber \\ 
r^{2}&\,=\,&R^{2} 
\, 
\left( {1+ 
\beta ^{2}} 
\right) +2 
\, 
\beta 
\, R 
\, R_{z}+2 
\, z 
\, 
\left( { 
\beta 
\, R 
\, + 
\, R_{z}} 
\right) +z^{2} 
\end{eqnarray} 
where $z \left( {t} 
\right)$ denotes the distance between the field 
point and the position of the charge at time $t$. Using spherical coordinates: 
\begin{equation} 
\vec {R}=R 
\, 
\sin \nolimits 
\theta 
\, 
\cos \nolimits 
\varphi 
\, 
\, 
\vec {i}+R 
\, 
\sin \nolimits 
\theta 
\, 
\sin \nolimits 
\varphi 
\, 
\, 
\vec {j}+R 
\, 
\cos \nolimits 
\theta 
\, 
\, 
\vec {k} 
\; , 
\; 
d^{3}R=R^{2} 
\, 
\sin \nolimits 
\theta 
\, dR 
\, d 
\theta 
\, d 
\varphi 
\end{equation} 
expression (36) becomes: 
\begin{eqnarray} 
A_{C2z}&=& 
\, -{{e 
\, v_{z}} \over {4 
\, 
\pi 
\, c}} 
\int \limits _{0}^{2 
\, 
\pi } 
\! 
\! 
\int \limits _{0}^{ 
\pi } 
\! 
\! 
\int \limits _{0}^{ 
\infty } 
\left[ {{{1} \over {r^{3}}}-{{3 
\, 
\, 
\left( {R 
\, 
\left( { 
\beta + 
\cos \nolimits 
\theta } 
\right) +z} 
\right) ^{2}} \over {r^{5}}}} 
\right] 
\, R 
\, 
\sin \nolimits 
\theta 
\, dR 
\, d 
\theta 
\, d 
\varphi \nonumber \\ 
r^{2}&\,=\,&R^{2} 
\, 
\left( {1+ 
\beta ^{2}+2 
\, 
\beta 
\, 
\cos \nolimits 
\theta } 
\right) +2 
\, R 
\, z 
\, 
\, 
\left( { 
\beta + 
\cos \nolimits 
\theta } 
\right) +z^{2} 
\end{eqnarray} 
The integration over $\varphi$ 
yields a factor of $2 
\, 
\pi$, since the integrand is independent of $ 
\varphi$. Upon indefinite integration over $R$ and $\theta$ one obtains: 
\begin{eqnarray} 
&A_{C2z}&= \\ 
&-&{{e 
\, v_{z}} \over {2 
\, c 
\, z}} 
\, {{R^{2} 
\, 
\left( {R+ 
\left( { 
\beta 
\, R+z} 
\right) 
\, 
\cos \nolimits 
\theta } 
\right) } \over { 
\left( { 
\beta 
\, R+z} 
\right) ^{2} 
\, \sqrt{R^{2} 
\, 
\left( {1+ 
\beta ^{2}+2 
\, 
\beta 
\, 
\cos \nolimits 
\theta } 
\right) +2 
\, R 
\, z 
\, 
\left( { 
\beta + 
\cos \nolimits 
\theta } 
\right) +z^{2}}}} \nonumber 
\end{eqnarray} 
The integral vanishes both at $R = 0$ and at $R = \infty$. 
It is singular at $R=z/ 
\left( {1- 
\beta } 
\right) 
\, , 
\, 
\, 
\theta = 
\pi$. Close 
to the singularity we expand it by substituting $R=z/ 
\left( {1- 
\beta } 
\right) -z 
\, 
\epsilon _{1} 
\, , 
\, 
\, 
\theta = 
\pi - 
\epsilon _{2}$ and obtain in lowest order: 
\begin{equation} 
A_{C2z}={{e 
\, v_{z}} \over {2 
\, c 
\, z}} 
\, 
\left( {1- 
\beta } 
\right) ^{2} 
\, 
\left( {{{ 
\epsilon _{1}} \over { 
\epsilon _{2}}}-{{ 
\left( {1- 
\beta } 
\right) 
\, 
\, 
\left( {5-9 
\, 
\beta } 
\right) } \over {4}} 
\, {{ 
\epsilon _{1}^{2}} \over { 
\epsilon _{2}}}} 
\right) 
\end{equation} 
Obviously, the integral assumes no definite value when we go to the 
limits $ 
\epsilon _{1}= 
\epsilon _{2}=0$ 
, as it does not converge absolutely. 

If we perform the same calculation on (29) we obtain in addition 
to undefined terms a diverging contribution: 
\begin{equation} 
E_{2z}= 
\, -{{e 
\, v_{z}^{2}} \over {2 
\, c^{2} 
\, z^{2}}} 
\, 
\, {{1- 
\beta } \over { 
\epsilon _{1}}} 
\, 
\, + 
\; {{0} \over {0}} 
\end{equation} 
which is also encountered when we calculate (31). 

From these results we must conclude that Maxwell's equations 
do not yield a physical solution for the fields of a moving particle 
in Coulomb gauge. Furthermore, the undefined fields as derived from 
(25), (27) and (28) by using (6) disagree definitely with the fields 
as given by (21) in Lorenz gauge which are well defined. Similar 
conclusions were reached by Onoochin [3] without evaluating the 
integral (33) explicitly. 

Jackson has attempted [4] to find an approximate 
`quasistatic' solution in Coulomb gauge which should 
be valid for velocities $v\ll c$ in a region very close 
to the particle where retardation may be neglected. We discuss this 
attempt in Appendix A and show that it also leads to an inconsistency. 

In a recent paper [6] Hnizdo has given a solution 
of (26) which is based on the gauge function as defined in equation 
(66) below in Section 7, and on the formal solution (67) for the 
gauge function. The second formal solution (70) is ignored in this 
consideration. Hnizdo arrives at a similar result as we found in (40), 
but he establishes uniqueness by applying a regularization procedure 
which can hardly be justified from a mathematical point of view. We 
discuss his approach in Appendix B. 

\vskip 30pt 

\section{Helmholtz's ansatz} 
Having obtained contradicting solutions in Lorenz gauge and 
in Coulomb gauge we infer that Maxwell's equations contain 
an inconsistency which does not permit to find a unique solution for 
the fields. In order to trace this problem, we employ Helmholtz's 
theorem which states that any vector field may be expressed as the 
sum of a rotational and an irrotational field. This was already used 
in the ansatz (6). Now we apply it to the electric field: 
\begin{equation} 
\vec {E}=rot 
\, 
\vec {U}- 
\nabla 
\phi 
\; 
\; , 
\; 
\; 
\; div 
\, 
\vec {U}=0 
\end{equation} 
Substituting this into (2) we obtain: 
\begin{equation} 
\Delta 
\vec {U}={{1} \over {c}} 
\, {{ 
\partial 
\vec {B}} \over { 
\partial t}} 
\end{equation} 
Taking the rotation of (4) and inserting (42) yields a second Poisson 
equation: 
\begin{equation} 
\Delta 
\, 
\left( { 
\vec {B}-{{1} \over {c}} 
\, {{ 
\partial 
\vec {U}} \over { 
\partial t}}} 
\right) 
\, = 
\, -{{4 
\, 
\pi } \over {c}} 
\, 
\nabla 
\rho 
\times 
\vec {v} 
\end{equation} 
Its solution for a point charge is: 
\begin{equation} 
\vec {B} 
\, ={{1} \over {c}} 
\, {{ 
\partial 
\vec {U}} \over { 
\partial t}} 
\, +{{e} \over {c}} 
\, 
\left( {{{ 
\vec {v} 
\times 
\vec {r}} \over {r^{3}}}} 
\right) 
\end{equation} 
Substituting this into (43) yields a wave equation for the vector 
potential of the electric field: 
\begin{equation} 
\Delta 
\vec {U}-{{1} \over {c^{2}}} 
\, {{ 
\partial ^{2} 
\vec {U}} \over { 
\partial t^{2}}}={{e} \over {c^{2}}} 
\, {{ 
\partial } \over { 
\partial t}} 
\, 
\left( {{{ 
\vec {v} 
\times 
\vec {r}} \over {r^{3}}}} 
\right) 
\end{equation} 
which has the retarded solution: 
\begin{eqnarray} 
\vec {U}\,= 
\, &-&{{e} \over {4 
\, 
\pi 
\, c^{2}}} 
\int \limits 
\! 
\! 
\int \limits 
\! 
\! 
\int \limits 
\left[ {{{ 
\partial } \over { 
\partial t'}} 
\, 
\left( {{{ 
\vec {v} 
\times 
\vec {r}} \over {r^{3}}}} 
\right) } 
\right] _{t'=t-{{R} \over {c}}}{{d^{3}s} \over {R}} \nonumber \\ 
\vec {r}&\,=\,& 
\vec {s}- 
\vec {x} 
\, ' 
\left( {t'} 
\right) 
\; , 
\; 
\;\; 
\vec {R}= 
\vec {s}- 
\vec {x} 
\end{eqnarray} 
For the magnetic field we obtain with (45): 
\begin{equation} 
\vec {B}= 
\, -{{e} \over {4 
\, 
\pi 
\, c^{3}}} 
\int \limits 
\! 
\! 
\int \limits 
\! 
\! 
\int \limits 
\left[ {{{ 
\partial ^{2}} \over { 
\partial t'^{2}}} 
\, 
\left( {{{ 
\vec {v} 
\times 
\vec {r}} \over {r^{3}}}} 
\right) } 
\right] _{t'=t-{{R} \over {c}}}{{d^{3}s} \over {R}}+{{e} \over {c}} 
\, 
\left( {{{ 
\vec {v} 
\times 
\vec {r}} \over {r^{3}}}} 
\right) 
\end{equation} 
and the electric field as derived from (47) with (42) becomes: 
\begin{eqnarray} 
\vec {E}= 
\, &-&{{e} \over {4 
\, 
\pi 
\, c^{2}}} 
\int \limits 
\! 
\! 
\int \limits 
\! 
\! 
\int \limits 
\left[ {{{ 
\vec {R}} \over {R^{3}}} 
\times 
\left( {{{R} \over {c}} 
\, {{ 
\partial ^{2}} \over { 
\partial t'^{2}}} 
\, 
\left( {{{ 
\vec {v} 
\times 
\vec {r}} \over {r^{3}}}} 
\right) +{{ 
\partial } \over { 
\partial t'}} 
\, 
\left( {{{ 
\vec {v} 
\times 
\vec {r}} \over {r^{3}}}} 
\right) } 
\right) } 
\right] _{t'=t-{{R} \over {c}}} \!\!\!\!\!\!\!d^{3}s \nonumber \\ 
&+&{{e 
\, 
\vec {r}} \over {r^{3}}} 
\end{eqnarray} 
where we have added the Coulomb field which results from insertion 
of (42) into (1). We note that neither of the expressions (48) and 
(49) agrees with the fields as calculated in Sections 2 and 3, 
because the fields derived from the Helmholtz ansatz (42) depend on 
the second time derivative of the velocity. Assuming a constant velocity 
of the particle one could also show that the integrals (48) and (49) 
actually diverge.

\vskip 30pt 
\section{Direct solution of the field equations} 

The two types of potential ansatz (6) and (42) resulted in 
different solutions for the fields. We, therefore, want to calculate 
the fields directly from (1 - 4) without using any potential 
ansatz. By elimination of the electric and the magnetic field, respectively, 
we find the wave equations: 
\begin{equation} 
\Delta 
\vec {B}-{{1} \over {c^{2}}} 
\, {{ 
\partial ^{2} 
\vec {B}} \over { 
\partial t^{2}}}= 
\, 
\! 
\, {{4 
\, 
\pi } \over {c}} 
\, 
\nabla 
\rho 
\times 
\vec {v} 
\end{equation} 
\begin{equation} 
\vec {E}-{{1} \over {c^{2}}} 
\, {{ 
\partial ^{2} 
\vec {E}} \over { 
\partial t^{2}}}= 
\, 
\! 
\, {{4 
\, 
\pi } \over {c^{2}}} 
\, {{ 
\partial } \over { 
\partial t}} 
\, 
\left( { 
\rho 
\, 
\vec {v}} 
\right) +4 
\, 
\pi 
\, 
\nabla 
\rho 
\end{equation} 
By applying the standard method of solving this kind of wave equation, 
as it was described in Section 2, it can be shown that the ensuing 
solution of (50) and (51) yields exactly the fields as given by (21). 
However, by deriving the hyperbolic equations (50) and (51) we have 
ignored the fact that Maxwell's equations are actually a mixture 
of hyperbolic \underline{and} elliptic equations which 
became very obvious in the previous Section. In order to take this 
into account we split the electric field into its irrotational and 
its rotational part: 
\begin{equation} 
\vec {E}= 
\vec {E}_{g}+ 
\vec {E}_{r} 
\end{equation} 
The rotational part does not enter equation (1). The irrotational 
part is just the quasistatic Coulomb field which does not propagate: 
\begin{equation} 
\vec {E}_{g}={{e 
\, 
\vec {r}} \over {r^{3}}} 
\; 
\; , 
\; 
\; 
\; 
\vec {r}= 
\vec {x}- 
\vec {x} 
\, ' 
\left( {t} 
\right) 
\end{equation} 
The rotational part obeys the wave equation: 
\begin{equation} 
\Delta 
\vec {E}_{r}-{{1} \over {c^{2}}} 
\, {{ 
\partial ^{2} 
\vec {E}_{r}} \over { 
\partial t^{2}}}= 
\, 
\! 
\, {{4 
\, 
\pi } \over {c^{2}}} 
\, {{ 
\partial } \over { 
\partial t}} 
\, 
\left( { 
\rho 
\, 
\vec {v}} 
\right) +{{ 
\, e} \over {c^{2}}} 
\, {{ 
\partial ^{2}} \over { 
\partial t^{2}}} 
\, 
\left( {{{ 
\vec {r}} \over {r^{3}}}} 
\right) 
\end{equation} 
which has the retarded solution: 
\begin{eqnarray} 
&\vec {E}_{r}& 
\, 
\, 
\left( { 
\vec {x} 
\, ,t} 
\right) 
\,= \\ 
&-& 
\int \limits 
\! 
\! 
\int \limits 
\! 
\! 
\int \limits 
\left[ {{{1} \over {c^{2}}} 
\, {{ 
\partial 
\, 
\left( { 
\rho 
\, 
\left( { 
\vec {s}, 
\, t'} 
\right) 
\, 
\vec {v} 
\, 
\left( {t'} 
\right) } 
\right) } \over { 
\partial t'}}+{{e} \over {4 
\, 
\pi 
\, c^{2}}} 
\, {{ 
\partial ^{2}} \over { 
\partial t'^{2}}} 
\, 
\left( {{{ 
\vec {r}} \over {r^{3}}}} 
\right) } 
\right] _{t'=t-{{ 
\left\vert { 
\vec {x}- 
\vec {s} 
\, } 
\right\vert } \over {c}}}{{ 
\, d^{3}s} \over { 
\left\vert { 
\vec {x}- 
\vec {s} 
\, 
\! 
\, } 
\right\vert }} \nonumber 
\end{eqnarray} 
Adding (53) one obtains the electric field as it was derived in Section 3 
in Coulomb gauge, whereas the retarded solution of (50) yields 
the magnetic field as it was derived in Section 2 in Lorenz gauge. 
It was shown in Section 3 that the integral (55) diverges. Hence, 
it does not represent a physical solution for the rotational part 
of the electric field. 

This analysis shows that the inconsistency inherent to Maxwell's 
equations is not an artefact produced by employing a potential ansatz. 
It seems to result from the mixture of hyperbolic and elliptic differential 
equations for the fields, as they were formulated by Maxwell. Only 
in Lorenz gauge the elliptic equations are completely removed so that 
there is seemingly agreement between the solutions of the hyperbolic 
equations (50) and (51) and the hyperbolic potential equations (10) 
and (11). 

\vskip 30pt 
\section{The inhomogeneous wave equations for a moving point 
source} 

The discrepancies encountered in Sections 3 - 5 are 
apparently related to the fact that Maxwell's set of equations 
mixes hyperbolic and elliptic structures so that unique solutions 
are not possible. In view of this finding it is somewhat surprising 
that in Lorenz gauge the elliptic features seem to be removed altogether 
so that equations (10) and (11) yield the unique solutions (18) and 
(19), provided the advanced solutions are suppressed on physical grounds. 
A closer look at the inhomogeneous wave equation (10) reveals, however, 
that the elliptic character is still there, but concealed in the 
inhomogeneity. If it is brought out, it turns out that the solution 
(18) cannot be considered as unique. 

In order to see this we employ a different method of solution 
than that used in Section 2. Due to the linearity of (10) one may 
split the potential into two contributions: 
\begin{equation} 
\phi = 
\phi _{0}+ 
\phi _{1} 
\end{equation} 
The wave equation may then be split into a Poisson equation: 
\begin{equation} 
\Delta 
\phi _{0}= 
\, -4 
\, 
\pi 
\, 
\rho 
\end{equation} 
and into another wave equation: 
\begin{equation} 
\Delta 
\phi _{1}-{{1} \over {c^{2}}} 
\, {{ 
\partial ^{2} 
\phi _{1}} \over { 
\partial t^{2}}}={{1} \over {c^{2}}} 
\, {{ 
\partial ^{2} 
\phi _{0}} \over { 
\partial t^{2}}} 
\end{equation} 
where the solution of the elliptic equation (57) enters as an extended 
source\footnote{The ansatz (56) was also used in [2], 
but it was erroneously assumed that $ 
\partial ^{2} 
\phi _{0}/ 
\partial t^{2}$ in (58) 
may be neglected.} Adding (57) and (58) the 
wave equation (10) is recovered. The retarded solution of (58) is: 
\begin{eqnarray} 
\phi _{1} 
\, 
\left( { 
\vec {x}, 
\, t} 
\right) &=& 
\, -{{e} \over {4 
\, 
\pi 
\, c^{2}}} 
\int \limits 
\! 
\! 
\int \limits 
\! 
\! 
\int \limits 
\left[ {{{ 
\partial ^{2}} \over { 
\partial t'^{2}}} 
\, 
\left( {{{1} \over {r}}} 
\right) } 
\right] _{t'=t-{{R} \over {c}}}{{d^{3}s} \over {R}} \nonumber \\ 
\vec {r}&=& 
\vec {s}- 
\vec {x} 
\, ' 
\left( {t'} 
\right) 
\; 
\; , 
\; 
\; 
\; 
\vec {R}= 
\vec {s}- 
\vec {x} 
\end{eqnarray} 
where we have substituted the instantaneous Coulomb potential (25) 
resulting from a solution of (57). Carrying out the differentiation 
in (59) we have: 
\begin{equation} 
\phi _{1} 
\, 
\left( { 
\vec {x}, 
\, t} 
\right) ={{e} \over {4 
\, 
\pi 
\, c^{2}}} 
\int \limits 
\! 
\! 
\int \limits 
\! 
\! 
\int \limits 
\left[ {{{1} \over {r^{3}}} 
\, 
\left( {v^{2}- 
\vec {r} 
\cdot {{d 
\vec {v}} \over {dt'}}} 
\right) -{{3 
\, 
\left( { 
\vec {r} 
\cdot 
\vec {v}} 
\right) ^{2}} \over {r^{5}}}} 
\right] _{t'=t-{{R} \over {c}}}{{d^{3}s} \over {R}} 
\end{equation} 
This integral depends on the acceleration which is not the case for 
the potential (18). At constant velocity the integral is very similar 
to (33) and we know from the calculation in Section 3 that it has 
no defined limiting value according to (40). 

Similar considerations apply to the inhomogeneous wave equation 
(11), the solution of which may formally be written as: 
\begin{equation} 
\vec {A} 
\, 
\, 
\left( { 
\vec {x}, 
\, t} 
\right) ={{e 
\, 
\vec {v}} \over {c 
\, r}} 
\, -{{e} \over {4 
\, 
\pi 
\, c^{3}}} 
\int \limits 
\! 
\! 
\int \limits 
\! 
\! 
\int \limits 
\left[ {{{ 
\partial ^{2}} \over { 
\partial t'^{2}}} 
\, 
\left( {{{ 
\vec {v}} \over {r}}} 
\right) } 
\right] _{t'=t-{{R} \over {c}}}{{d^{3}s} \over {R}} 
\end{equation} 
by applying the same method of splitting the vector potential into 
two parts. If result (61) is substituted into (6), the fields would 
depend on the third derivative of the velocity, which is not the case 
according to (21), so that (61) is incompatible with (19). 

We must conclude then that the potentials (56) and (61) disagree 
with the Li\'enard-Wiechert potentials (18) and (19) which turn out 
not to be a unique solution of the inhomogeneous wave equations (10) 
and (11), even if the advanced solutions are suppressed. As a matter 
of fact, equations (10) and (11) have no physical solution judged 
from our results (60) and (61) which do not admit a defined limiting 
value. 

There is a direct way of showing that the Li\'enard-Wiechert 
result, which leads to the fields (21), satisfies only Maxwell's 
homogeneous equations, but the inhomogeneities are not taken 
into account properly. Let us consider Green's first identity: 
\begin{equation} 
\int \limits 
\! 
\! 
\int \limits 
\phi 
\, 
\nabla 
\phi 
\cdot d^{2} 
\vec {x}= 
\int \limits 
\! 
\! 
\int \limits 
\! 
\! 
\int \limits 
\left( { 
\phi 
\, 
\Delta 
\phi + 
\left\vert { 
\nabla 
\phi } 
\right\vert ^{2}} 
\right) 
\, d^{3}x 
\end{equation} 
The surface integral on the left-hand-side vanishes at infinity both 
for the Coulomb and for the Lorenz potential so that the volume integral 
over all space on the right-hand-side must vanish as well. Substituting 
the Coulomb potential (25) together with (23) one obtains an integral 
equation which must be satisfied by the charge density: 
\begin{equation} 
4 
\, 
\pi 
\, e 
\, 
\! 
\! 
\int \limits _{0}^{ 
\infty } 
\left( {-{{4 
\, 
\pi } \over {r}} 
\, 
\rho 
\, 
\left( {r} 
\right) +{{e} \over {r^{4}}}} 
\right) 
\, r^{2} 
\, dr=0 
\end{equation} 
where spherical coordinates centered around the position of the charge 
were used. If one inserts the Li\'enard-Wiechert potential (18) together 
with (10) into (62), one obtains: 
\begin{equation} 
2 
\, 
\pi 
\int \limits _{0}^{ 
\infty } 
\! 
\! 
\int \limits _{0}^{ 
\pi } 
\left( {-4 
\, 
\pi 
\, 
\phi _{LW} 
\, 
\rho 
\, 
\left( {r} 
\right) +{{ 
\phi _{LW}} \over {c^{2}}} 
\, {{ 
\partial ^{2} 
\phi _{LW}} \over { 
\partial t^{2}}}+ 
\left\vert { 
\nabla 
\phi _{LW}} 
\right\vert ^{2}} 
\right) 
\, 
\sin \nolimits 
\theta 
\, d 
\theta 
\, r^{2} 
\, dr=0 
\end{equation} 
This integral equation depends now on the velocity which may be easily 
verified by choosing a constant velocity so that (18) yields: 
\begin{equation} 
\phi _{LW}={{e} \over { 
\left[ { 
\left( { 
\vec {v} 
\cdot 
\vec {r}/c} 
\right) ^{2}+ 
\left( {1-v^{2}/c^{2}} 
\right) 
\, r^{2}} 
\right] ^{{{1} \over {2}}}}} 
\; 
\; , 
\; 
\; 
\; 
\vec {r}= 
\vec {x}- 
\left( { 
\vec {x}_{0}+ 
\vec {v} 
\, t} 
\right) 
\end{equation} 
Both integral equations (63) and (64) cannot be satisfied by the same 
function $ 
\rho 
\, 
\left( {r} 
\right)$, unless the `shape' of 
the point charge would depend on the velocity, as suggested by Onoochin 
in Reference [3].

\vskip 30pt 
\section{Transformation of the Lorenz potentials into Coulomb 
potentials} 

From the results obtained in Sections 2 - 6 it should 
be evident by now that a unique gauge transformation, which transforms 
the Lorenz potentials of a point source into the corresponding Coulomb 
potentials, cannot exist. We finally want to show this explicitly. 
The gauge transformation is effected by a generating function $\chi$: 
\begin{equation} 
\phi _{C}= 
\phi _{L}-{{1} \over {c}} 
\, {{ 
\partial 
\chi 
\, 
\left( { 
\vec {x}, 
\, t} 
\right) } \over { 
\partial t}} 
\; 
\; 
\; , 
\; 
\; 
\; 
\; 
\vec {A}_{C}= 
\vec {A}_{L}+ 
\nabla 
\chi 
\, 
\left( { 
\vec {x}, 
\, t} 
\right) 
\end{equation} 
By integrating the first relation over time one obtains immediately: 
\begin{equation} 
\chi 
\, 
\left( { 
\vec {x}, 
\, t} 
\right) = 
\int \limits _{t_{0}}^{t}c 
\, 
\, 
\left( { 
\phi _{L}- 
\phi _{C}} 
\right) 
\, 
\, dt+ 
\chi _{0} 
\, 
\left( { 
\vec {x}} 
\right) 
\end{equation} 
where the Lorenz potential is given by (18) and the Coulomb potential 
by (25). The gauge function must also satisfy the Poisson equation: 
\begin{equation} 
\Delta 
\chi 
\, 
\left( { 
\vec {x}, 
\, t} 
\right) =div 
\, 
\vec {A}_{C}-div 
\, 
\vec {A}_{L} 
\end{equation} 
which follows from the second relation in (66). Equation (68) has 
the instantaneous solution: 
\begin{equation} 
\chi 
\, 
\left( { 
\vec {x}, 
\, t} 
\right) = 
\, -{{1} \over {4 
\, 
\pi 
\, c}} 
\int \limits 
\! 
\! 
\int \limits 
\! 
\! 
\int \limits {{ 
\partial 
\phi _{L} 
\left( { 
\vec {s}, 
\, t} 
\right) } \over { 
\partial t}} 
\, {{d^{3}s} \over { 
\left\vert { 
\vec {s}- 
\vec {x}} 
\right\vert }} 
\end{equation} 
where we have substituted the conditions (9) and (22) into (68). If 
we insert the Lorenz potential as given by (18) and apply the first 
differentiation rule in (20), we find: 
\begin{eqnarray} 
\chi 
\, 
\left( { 
\vec {x}, 
\, t} 
\right) \,= 
\, &-&{{e} \over {4 
\, 
\pi 
\, c}} 
\int \limits 
\! 
\! 
\int \limits 
\! 
\! 
\int \limits {{1} \over { 
\lambda ^{3} 
\, R^{3}}} 
\, 
\left[ { 
\vec {R} 
\cdot 
\vec {v}+{{R} \over {c}} 
\, 
\left( { 
\vec {R} 
\cdot {{d 
\vec {v}} \over {dt'}}-v^{2}} 
\right) } 
\right] 
\, {{d^{3}s} \over { 
\left\vert { 
\vec {s}- 
\vec {x}} 
\right\vert }} \nonumber \\ 
\vec {R}&\,=\,& 
\vec {s}- 
\vec {x} 
\, ' 
\left( {t'} 
\right) 
\; 
\; , 
\; 
\; 
\; t'=t-{{ 
\left\vert { 
\vec {s}- 
\vec {x}} 
\right\vert } \over {c}} 
\end{eqnarray} 
As this expression depends not only on the velocity, but also on the 
acceleration, it is not compatible with expression (67). Furthermore, 
the integral (70) has no unique limiting value, but depends on the 
chosen integration variables. In order to see this we assume a constant 
velocity. The Li\'enard-Wiechert potential (18) becomes in this case: 
\begin{equation} 
\phi _{L} 
\left( { 
\vec {x}, 
\, t} 
\right) ={{e} \over { 
\left[ { 
\left( { 
\vec {v} 
\cdot 
\vec {r}/c} 
\right) ^{2}+ 
\left( {1-v^{2}/c^{2}} 
\right) 
\, r^{2}} 
\right] ^{{{1} \over {2}}}}} 
\; 
\; 
\; , 
\; 
\; 
\; 
\; 
\vec {r}= 
\vec {x}- 
\vec {x} 
\, ' 
\left( {t} 
\right) 
\end{equation} 
Substitution into (69) yields: 
\begin{equation} 
\chi 
\, 
\left( { 
\vec {x}, 
\, t} 
\right) = 
\, -{{e} \over {4 
\, 
\pi 
\, c}} 
\int \limits 
\! 
\! 
\int \limits 
\! 
\! 
\int \limits {{ 
\vec {v} 
\cdot 
\vec {s}} \over { 
\left[ { 
\left( { 
\vec {v} 
\cdot 
\vec {s}/c} 
\right) ^{2}+ 
\left( {1-v^{2}/c^{2}} 
\right) 
\, s^{2}} 
\right] ^{{{3} \over {2}}}}} 
\, {{d^{3}s} \over { 
\left\vert { 
\vec {r}- 
\vec {s}} 
\right\vert }} 
\end{equation} 
where we have chosen a coordinate system with its origin at the position 
$ 
\vec {x} 
\, ' 
\left( {t} 
\right)$ 
of the charge. If we change to the integration 
variable $ 
\vec {s} 
\, '= 
\vec {s}- 
\vec {r}$, which is equivalent to shifting the 
origin of the coordinate system to the field point $ 
\vec {x}$, we obtain instead: 
\begin{eqnarray} 
&\chi& 
\, 
\left( { 
\vec {x}, 
\, t} 
\right) \,= \\ 
&-&{{e} \over {4 
\, 
\pi 
\, c}} 
\int \limits 
\! 
\! 
\int \limits 
\! 
\! 
\int \limits {{ 
\vec {v} 
\cdot 
\left( { 
\vec {s} 
\, '+ 
\vec {r}} 
\right) } \over { 
\left[ { 
\left( { 
\vec {v} 
\cdot 
\left( { 
\vec {s} 
\, '+ 
\vec {r}} 
\right) /c} 
\right) ^{2}+ 
\left( {1-v^{2}/c^{2}} 
\right) 
\, 
\left( {s'^{2}+ 
\vec {r}^{ 
\, 2}-2 
\, 
\vec {s} 
\, ' 
\cdot 
\vec {r}} 
\right) } 
\right] ^{{{3} \over {2}}}}} 
\, {{d^{3}s'} \over {s'}} \nonumber 
\end{eqnarray} 
Apart from a common logarithmic singularity at infinity the integrals 
(72) and (73) are conditionally convergent and assume different limiting values 
which was verified by calculating them in spherical coordinates. Numerical 
calculations of (72) and (73) in cylindrical coordinates yield still 
different limiting values. None of these results agrees with (67). 

In accordance with the previous conclusions reached in this paper we 
infer that no unique function $\chi$ exists which would 
transform the Lorenz potentials of a point source into the corresponding 
Coulomb potentials. Hence, the principle of `gauge invariance' 
is not applicable to classical electrodynamics.

\vskip 30pt 
\section{Discussion} 

The nature of the inconsistencies encountered in Sections 2 - 5 is 
apparently connected to the feature of Maxwell's 
equations of mixing elliptic and hyperbolic structures. Even if the 
equations are reduced to wave equations like (10) and (11) or (50) 
and (51), the elliptic character is still there in form of the inhomogeneity 
and may be made visible by the method of solution employed in Section 
6. There we were compelled to conclude that the inhomogeneous wave 
equation does not have, as a matter of fact, a unique solution, or even leads 
to unphysical diverging solutions, at least in case of a moving point 
source. In principle, it is well known that the inhomogeneous wave 
equation has an infinite manifold of solutions, but it is generally 
believed that suppression of the advanced solutions reduces it to 
a physical solution, the properties of which are uniquely determined 
by the behaviour of the source. According to our analysis in Section 
6 we must maintain, however, that the inhomogeneous wave equations 
do not correctly describe the physical process of wave production 
by a moving point source. 

Our result is not too surprising, if we realize that the inhomogeneous 
wave equations (50) and (51) relate the measurable fields at a certain 
location with temporal changes in a remote source at the {\em same} time. 
Although both the fields and 
the sources in equations (1 - 4) were differentiated at time $t$ when, 
e.g., the electric field was eliminated 
to obtain (50), the retarded solutions (21) require that the differentiation 
of the sources is dated back to the earlier time $t-R/c$. However, 
the source may be an extinguished star the 
light of which we see only now at time $t$. It makes 
no sense to differentiate a source not existing any more which has 
no influence whatsoever on the light we see after a billion years. 
When we differentiate Maxwell's equations now to obtain the 
wave equations for the fields, we treat the temporal changes in the 
sources as if they would happen now at time $t$. In the retarded solutions, 
however, we date back the 
change in the sources to a remote past. This procedure is inconsistent, 
but inescapable due to the structure of Maxwell's equations. 

If the same procedure would be applied to acoustic waves, one 
would encounter similar inconsistencies. Instead, from the hydrodynamic equations one derives linearized homogeneous wave equations for 
the pressure and the fluid velocity. These are solved by imposing 
suitable boundary conditions which are determined, e.g., by the oscillating 
membrane of a loudspeaker. Maxwell's equations, however, lead 
to inhomogeneous wave equations which connect the {\em travelling} 
fields with the source at the {\em same} time, a {\it contradictio in 
adjecto}. This became quite obvious in Section 5 where equation 
(51) predicts that the {\em total} electric field 
is produced in a point-like region and travels within a finite time 
to the field point where an observer may be placed. On the other hand, 
equation (53) predicts that {\em part} of the 
field has already arrived there instantaneously, as soon as any change 
in the source occurred. This inconsistency cannot be resolved without 
altering equations (1 - 4). 

It is well possible that Maxwell was fully aware of this problem, 
because in his `Treatise' [7] he did {\em not} 
derive an inhomogeneous wave equation. 
He used the Coulomb gauge and derived equation (24). Then he argued 
that in the `ether' there does not exist any current or free 
charge. This way he was left with: 
\begin{equation} 
\Delta 
\vec {A}-{{1} \over {c^{2}}} 
\, {{ 
\partial ^{2} 
\vec {A}} \over { 
\partial t^{2}}}= 
\, {{1} \over {c}} 
\, {{ 
\partial } \over { 
\partial t}} 
\nabla 
\phi 
\; 
\; , 
\; 
\; 
\; 
\; 
\Delta 
\phi =0 \nonumber 
\end{equation} 
Now he committed a formal inconsistency by concluding that the vanishing 
of the Laplacian of the scalar potential justifies to omit also the 
gradient of the potential. This is, of course, not true close to a 
charged body. In other words, he omitted part of the `displacement 
current', which was invented by him in the first place, in 
order to `derive' a homogeneous wave equation for the 
vector potential. At last he suggested that this equation should be 
solved by imposing Cauchy-type boundary conditions on $ 
\vec {A}$ and $ 
\partial 
\vec {A}/ 
\partial t$. The result would be a travelling 
vector wave from which the measurable fields could be derived with 
(6) at any place and time where the wave has arrived. In his last Chapter XXIII (Article 862) Maxwell discusses Riemann's inhomogeneous wave equation which is formally the same as Lorenz's equation (10). Taking reference to Clausius he states that Riemann's formula is not in agreement with ``the known laws of electrodynamics''. Maxwell's method of using only a homogeneous wave equation to describe electromagnetic waves is still in practical use, when the radiation emitted by an antenna is calculated. Only the homogeneous wave equation is used together with plausible boundary conditions resulting from 
a physics which goes beyond Maxwell's equations. 

In parentheses we remark that Einstein gave his famous paper 
of 1905 the title: ``Zur Elektrodynamik bewegter K\"orper'', 
but the `moving bodies' which carry charges and currents 
are not treated in his analysis. He deals only with Maxwell's 
{\em homogeneous} equations which do not lead 
to contradictions. The {\em instantaneous} Coulomb 
potential is left out of the consideration. 

For slowly varying fields Maxwell's equations do make 
sense. They describe correctly the phenomenon of induction in transformers 
where only the instantaneous fields come into play and where the displacement 
current is negligible. When a condenser is 
charged up, the displacement current must be allowed for, but then it is only the instantaneous Coulomb field which 
matters in practice. A quasistatic `near field' theory 
can be carried through satisfactorily, but amalgamating it with wave 
phenomena, in the way Maxwell has tried it, leads to the contradictions 
which we have demonstrated. These were also recognized by Dunn [8], but 
not worked out in detail. 

\vskip 30pt 

\noindent 
\textbf{Appendix A} 

\noindent The wave equation (26) for the second part of the `Coulomb' 
vector potential: 
\setcounter{equation}{0} 
\renewcommand{\theequation}{A.\arabic{equation}} 
\begin{equation} 
\Delta 
\vec {A}_{C2}-{{1} \over {c^{2}}} 
\, {{ 
\partial ^{2} 
\vec {A}_{C2}} \over { 
\partial t^{2}}}={{e} \over {c}} 
\, 
\left( {{{ 
\vec {v}} \over {r^{3}}}-{{3 
\, 
\vec {r} 
\, 
\, 
\left( { 
\vec {v} 
\cdot 
\vec {r}} 
\right) } \over {r^{5}}}} 
\right) 
\end{equation} 
may be taken as a Poisson equation which has the formal solution: 
\begin{eqnarray} 
\vec {A}_{C2}\,&=&\,-{{e} \over {4 
\, 
\pi 
\, c}} 
\int \limits 
\! 
\! 
\int \limits 
\! 
\! 
\int \limits 
\left( {{{ 
\vec {v}} \over {r^{3}}}-{{3 
\, 
\vec {r} 
\, 
\, 
\left( { 
\vec {v} 
\cdot 
\vec {r}} 
\right) } \over {r^{5}}}} 
\right) {{d^{3}s} \over { 
\left\vert { 
\vec {s}- 
\vec {x}} 
\right\vert }} \nonumber \\ 
&-&\,{{1} \over {4 
\, 
\pi 
\, c^{2}}} 
\int \limits 
\! 
\! 
\int \limits 
\! 
\! 
\int \limits {{ 
\partial ^{2} 
\vec {A}_{C2}} \over { 
\partial t^{2}}} 
\, {{d^{3}s} \over { 
\left\vert { 
\vec {s}- 
\vec {x}} 
\right\vert }} \quad ,\qquad 
\vec {r}= 
\vec {s}- 
\vec {x} 
\, ' 
\left( {t} 
\right) 
\end{eqnarray} 
Close to the charge and at small velocity $v\ll c$ 
the second term may be expected to be negligibly small so that the first 
integral could be considered as an approximate solution of (A1) with 
limited applicability. 

This attempt to obtain a `quasistatic' solution 
was pursued by Jackson [4] in order to find the interaction 
Lagrangian between two particles moving at nonrelativistic velocities. 
He chose a coordinate system centered at $ 
\vec {x} 
\, '=0$ and 
performed a partial integration: 
\begin{eqnarray} 
-&{{4 
\, 
\pi 
\, c} \over {e}} 
\, 
\vec {A}_{C2}&\,= 
\int \limits 
\! 
\! 
\int \limits 
\! 
\! 
\int \limits 
\, {{ 
\partial } \over { 
\partial 
\vec {s}}} 
\, 
\left( {{{ 
\vec {v} 
\cdot 
\vec {s}} \over {s^{3}}}} 
\right) 
\, {{d^{3}s} \over { 
\left\vert { 
\vec {s}- 
\vec {x}} 
\right\vert }} \\ 
&=&\!\!\!\!\!\! 
- 
\int \limits 
\! 
\! 
\int \limits 
\! 
\! 
\int \limits 
\, 
\, 
\left( {{{ 
\vec {v} 
\cdot 
\vec {s}} \over {s^{3}}}} 
\right) 
\, {{ 
\partial } \over { 
\partial 
\vec {s}}} 
\, 
\left( {{{1} \over { 
\left\vert { 
\vec {s}- 
\vec {x}} 
\right\vert }}} 
\right) 
\, d^{3}s 
={{ 
\partial } \over { 
\partial 
\vec {x}}} 
\int \limits 
\! 
\! 
\int \limits 
\! 
\! 
\int \limits 
\, 
\, {{ 
\vec {v} 
\cdot 
\vec {s}} \over {s^{3}}} 
\, {{d^{3}s} \over { 
\left\vert { 
\vec {s}- 
\vec {x}} 
\right\vert }} \nonumber 
\end{eqnarray} 
Now the integration was straightforward and yielded: 
\begin{equation} 
\vec {A}_{C2}= 
\, -{{e} \over {c}} 
\, {{ 
\partial } \over { 
\partial 
\vec {x}}} 
\, 
\left( {{{ 
\vec {v} 
\cdot 
\vec {x}} \over {2 
\, 
\left\vert { 
\vec {x}} 
\right\vert }}} 
\right) ={{e} \over {2 
\, c}} 
\, 
\left( {-{{ 
\vec {v}} \over { 
\left\vert { 
\vec {x}} 
\right\vert }}+{{ 
\vec {x} 
\, 
\, 
\left( { 
\vec {v} 
\cdot 
\vec {x}} 
\right) } \over { 
\left\vert { 
\vec {x}} 
\right\vert ^{3}}}} 
\right) 
\end{equation} 
In order to obtain the total vector potential, the unretarded contribution 
from expression (27) must be added: 
\begin{equation} 
\vec {A}_{C}={{e} \over {2 
\, c}} 
\, 
\left( {{{ 
\vec {v}} \over { 
\left\vert { 
\vec {x}} 
\right\vert }}+{{ 
\vec {x} 
\, 
\, 
\left( { 
\vec {v} 
\cdot 
\vec {x}} 
\right) } \over { 
\left\vert { 
\vec {x}} 
\right\vert ^{3}}}} 
\right) 
\end{equation} 

It turns out, however, that the solution (A.4) does not satisfy 
the Poisson equation from which it was calculated. Substituting (A.4) 
into the left-hand-side of (A.1) and ignoring the second time derivative 
yields: 
\begin{equation} 
-{{e 
\, 
\vec {v}} \over {2 
\, c}} 
\, 
\Delta 
\, 
\left( {{{1} \over { 
\left\vert { 
\vec {x}} 
\right\vert }}} 
\right) +{{e} \over {c}} 
\, 
\left( {{{ 
\vec {v}} \over { 
\left\vert { 
\vec {x}} 
\right\vert ^{3}}}-{{3 
\, 
\vec {x} 
\, 
\, 
\left( { 
\vec {v} 
\cdot 
\vec {x}} 
\right) } \over { 
\left\vert { 
\vec {x}} 
\right\vert ^{5}}}} 
\right) 
\end{equation} 
The distance vector $\vec {x}$ pointing from the origin 
to the field point may be identified with $\vec {r}$, since 
the position of the charge was assumed at the origin. The second term 
of (A.6) equals the right-hand-side of (A.1), but the first term yields 
a $\delta$ - function: 
\begin{equation} 
-{{e 
\, 
\vec {v}} \over {2 
\, c}} 
\, 
\Delta 
\, 
\left( {{{1} \over { 
\left\vert { 
\vec {x}} 
\right\vert }}} 
\right) ={{2 
\, 
\pi 
\, e 
\, 
\vec {v}} \over {c}} 
\, 
\delta 
\, 
\left( { 
\vec {x}} 
\right) ={{2 
\, 
\pi 
\, 
\rho 
\, 
\left( { 
\vec {x}} 
\right) 
\, 
\vec {v}} \over {c}} 
\end{equation} 
which remains unaccounted for in (A.1). 

The reason for the discrepancy is that the first integral in 
(A.2) does not converge absolutely, as we have shown in Section 3. 
Consequently, the operations of partial integration as well as interchanging 
the sequence of differentiation and integration in (A.3) are not permitted 
and lead to an incorrect result. The first integral in (A.2) has, in 
fact, no defined limiting value as is obvious from expression (40). 

In a private communication Professor Jackson explained how 
he could obtain (A5) from an expansion procedure applied on the 
Li\'enard-Wiechert 
potential - which is based on the Lorenz gauge - and 
that this was actually being done by Darwin who derived (A.5) in 1920 
[5].

\vskip 30pt 
\noindent 
\textbf{Appendix B} 

\noindent 
Hnizdo's article {\em Potentials of a uniformly 
moving point charge in the Coulomb gauge} [6] 
was apparently motivated by Onoochin's objections against `mainstream' 
electrodynamics as published in Ref. [3]. Onoochin 
reached conclusions which are similar to those arrived at in the present 
paper, namely that the electrodynamic field depends on the choice 
of the gauge. He speculates that the `shape' of the 
electron could depend on its velocity, an idea which was already pursued 
(unsuccessfully) by Lorentz. Hnizdo tries to resolve the problem by 
using a certain regularization procedure when the integral (33) is 
evaluated. 

Both authors do not emphasize that this type of integral is 
conditionally convergent. This property may lead to the known fact that the value of the integral depends on the sequence of integration, on the chosen 
coordinate system, or on the position of the origin. A typical example 
is given in Bronstein-Semendjajew's {\em Taschenbuch 
der Mathematik}, Verlag Harri Deutsch, Frankfurt, on 
page 347: 
\setcounter{equation}{0} 
\renewcommand{\theequation}{B.\arabic{equation}} 
\begin{equation} 
\int \limits _{0}^{1} 
\! 
\! 
\int \limits _{0}^{1}{{y^{2}-x^{2}} \over { 
\left( {x^{2}+y^{2}} 
\right) ^{2}}} 
\, dx 
\, dy={{ 
\pi } \over {4}} 
\; 
\hbox { 
\relax or:} 
\; = 
\; -{{ 
\pi } \over {4}} 
\end{equation} 
The result depends on whether one integrates first over $x$ 
and then over $y$, or vice 
versa. Although one obtains a finite result in both cases, its value 
is not unique. 

Hnizdo uses the formal solution (67) to obtain his formulas 
(16 - 18) for the difference between the vector potentials in 
Coulomb and in Lorenz gauge. He claims that these formulae do not 
have singularities, but this is actually not true. The $y$-component 
for the difference is, e.g.: 
\begin{equation} 
A_{Cy}-A_{Ly}= 
\, -{{c} \over {v}} 
\; {{y 
\, 
\left( {x-v 
\, t} 
\right) } \over {y^{2}+z^{2}}} 
\; 
\left( {V_{C}-V_{L}} 
\right) 
\end{equation} 
If this expression is expanded around the point \\ $ 
\left( {x=v 
\, t+ 
\epsilon _{1} 
\, , 
\> y=0+ 
\epsilon _{2} 
\, , 
\> z=0+ 
\epsilon _{3}} 
\right)$, one obtains: 
\begin{equation} 
A_{Cy}-A_{Ly}= 
\, -{{c} \over {v}} 
\; {{ 
\epsilon _{2} 
\, 
\epsilon _{1}} \over { 
\epsilon ^{2}_{2}+ 
\epsilon ^{2}_{3}}} 
\; 
\left( {V_{C}-V_{L}} 
\right) 
\end{equation} 
This result is similar to our result (40) where we concluded that 
no limiting value exists. In fact, (B.3) can assume any value between 
zero and infinity depending on the way how one approaches the limits 
$\epsilon _{1} 
\rightarrow 0, 
\, 
\epsilon _{2} 
\rightarrow 0, 
\, 
\epsilon _{3} 
\rightarrow 0$. 

Hnizdo avoids the ambiguity by using a regularization procedure 
as defined in his equation (32). It amounts to using spherical coordinates 
centered around the charge point. This prescription is, however, arbitrary 
and lacks a rigorous justification. If one deviates from 
it, one finds non-uniqueness as demonstrated in our comparison of 
equations (72) and (73). It should be noted that Hnizdo's regularization 
would not work, when applied to the expression (70) for the gauge 
function which is incompatible with (67). This demonstrates again 
that Maxwell's equations do not have a unique solution (for 
moving point charges) representing a measurable well defined 
field. 

\vskip 30pt 
\noindent 
\textbf{Acknowledgments} 

\noindent The author is deeply indebted to Dr. O. Kardaun for extensive 
and stimulating discussions of the subject. In particular, he pointed 
out that the vector potential in Coulomb gauge leads to a conditionally
convergent integral. He also pointed to the mixed character of Maxwell's 
equations involving elliptic and hyperbolic equations. 

A communication with Professor J. D. Jackson in October 2001 
was very useful for clarifying the issue dealt with in Appendix A. 

Critical previous comments by Professor D. Pfirsch and Dr. 
R. Gruber helped to formulate the paper more concisely than originally 
conceived. Professor Pfirsch's written comments, in particular, 
alluded to the possibility that a gauge transformation between Lorenz 
and Coulomb potentials might not exist, as discussed in Section 7. 

\vskip 30pt

\begin{eref} 

\bibitem{} J. D. Jackson, L. B. Okun, Reviews of Modern Physics, {\bf 73}, 
663, (2001). 

\bibitem{} Our calculation follows the methods applied in: \\ 
R. Becker, F. Sauter, {\em Theorie der Elektrizit\"at}, 
B. G. Teubner Verlagsgesellschaft, Stuttgart (1962), 
\S\S \;66, 69. 

\bibitem{} V. V. Onoochin, Annales de la Fondation Louis de Broglie, 
{\bf 27}, 163, (2002). 

\bibitem{} D. Jackson, {\em Classical Electrodynamics}, Third Edition, 
John Wiley \& Sons, Inc., New York (1999), Chapter 12.6: derivation 
of equation (12.80). 

\bibitem{} C. G. Darwin, Phil. Mag. ser. 6, {\bf 39}, 537, (1920). 

\bibitem{}V. Hnizdo, European Journal of Physics, {\bf 25}, 351, (2004). 

\bibitem{} J. C. Maxwell, {\em A Treatise on Electricity and Magnetism}, 
Dover Publications, Inc., New York (1954), Vol. 2, Articles 783, 784. 

\bibitem{}D. A. Dunn, {\em Models of Particles and Moving Media}, Academic 
Press, New York and London (1971), page 118ff. 

\end{eref} 

\man{30 ao\^ut 2004}

\end{document}